\documentstyle[epsf]{elsart}

\def\lta{\mathrel{\vcenter{\vbox{\offinterlineskip \hbox{$<$}
     \vskip 0.2 pt \hbox{$\sim$}}}}}
\def\tot{{\rm tot}}

\begin{document}

\flushright{MIT-CTP-2948, astro-ph/0002156}

\begin{frontmatter}

\title{Inflation and Eternal Inflation}

\author{Alan H. Guth}

\address{Center for Theoretical Physics, Laboratory for Nuclear
Science, and \\
Department of Physics, Massachusetts Institute of Technology, \\
Cambridge, Massachusetts 02139, USA\protect\footnote{Present
address.} \\ and \\
Isaac Newton Institute for Mathematical Sciences, \\
Clarkson Road, Cambridge CB3 0EH, UK} 

\begin{abstract}
The basic workings of inflationary models are summarized, along
with the arguments that strongly suggest that our universe is the
product of inflation.  The mechanisms that lead to eternal
inflation in both new and chaotic models are described.  Although
the infinity of pocket universes produced by eternal inflation
are unobservable, it is argued that eternal inflation has real
consequences in terms of the way that predictions are extracted
from theoretical models.  The ambiguities in defining
probabilities in eternally inflating spacetimes are reviewed,
with emphasis on the youngness paradox that results from a
synchronous gauge regularization technique.  Vilenkin's proposal
for avoiding these problems is also discussed.
\end{abstract}

\thanks{This work is supported in part by funds provided by the
U.S. Department of Energy (D.O.E.) under cooperative research
agreement \#DF-FC02-94ER40818, and in part by funds provided by
NM Rothschild \& Sons Ltd and by the EPSRC.}

\end{frontmatter}

\setcounter{footnote}{0}

\section{Introduction}

     There are many fascinating issues associated with eternal
inflation, so I can think of no subject more appropriate to
discuss in a volume commemorating David Schramm.  The shock of
Dave's untimely death showed that even the most vibrant of human
lives is not eternal, but his continued influence on our entire
field proves that in many ways David Schramm is truly eternal. 
Dave is largely responsible for creating the interface between
particle physics and cosmology, and is very much responsible for
cementing together the community in which this interface
developed.  His warmth, his enthusiasm, and the efforts that he
made to welcome young scientists to the field have strengthened
our community in a way that will not be forgotten. 

I will begin by summarizing the basics of inflation, including a
discussion of how inflation works, and why many of us believe
that our universe almost certainly evolved through some form of
inflation.  This material is not new, but I think it should
certainly be included in any volume that attempts to summarize
the important advances that Dave helped to develop and promote. 
Then I will move on to discuss eternal inflation, attempting to
emphasize that this topic has important implications, and raises
important questions, which should not be dismissed as being
metaphysical.

\section{How Does Inflation Work?}

The key property of the laws of physics that makes inflation
possible is the existence of states of matter that have a high
energy density which cannot be rapidly lowered.  In the original
version of the inflationary theory \cite{Guth1}, the proposed
state was a scalar field in a local minimum of its potential
energy function.  A similar proposal was advanced by Starobinsky
\cite{Starobinsky}, in which the high energy density state was
achieved by curved space corrections to the energy-momentum
tensor of a scalar field.  The scalar field state employed in the
original version of inflation is called a {\it false vacuum},
since the state temporarily acts as if it were the state of
lowest possible energy density.  Classically this state would be
completely stable, because there would be no energy available to
allow the scalar field to cross the potential energy barrier that
separates it from states of lower energy.  Quantum mechanically,
however, the state would decay by tunneling \cite{Coleman}. 
Initially it was hoped that this tunneling process could
successfully end inflation, but it was soon found that the
randomness of false vacuum decay would produce catastrophically
large inhomogeneities.  These problems were summarized in
Ref.~\cite{Guth1}, and described more fully by Hawking, Moss, and
Stewart \cite{HMS} and by Guth and Weinberg
\cite{GuthWeinberg}. 

\begin{figure}
\epsfxsize=201pt
\centerline{\epsfbox{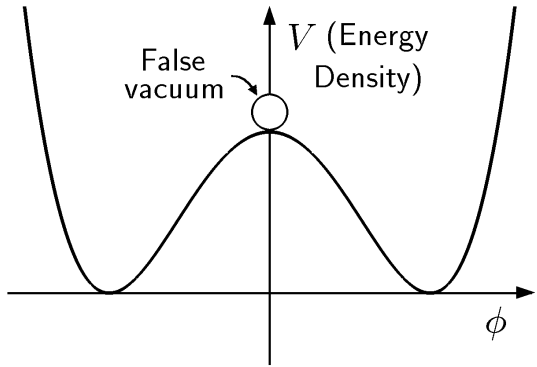}}
\caption{Generic form of the potential for the new inflationary
scenario.}
\label{newinf}
\end{figure}

This ``graceful exit'' problem was solved by the invention of the
new inflationary universe model by Linde \cite{Linde1} and by
Albrecht and Steinhardt \cite{Albrecht-Steinhardt1}.  New
inflation achieved all the successes that had been hoped for in
the context of the original version.  In this theory inflation is
driven by a scalar field perched on a plateau of the potential
energy diagram, as shown in Fig.~\ref{newinf}.  Such a scalar
field is generically called the {\it inflaton}.  If the plateau
is flat enough, such a state can be stable enough for successful
inflation. Soon afterwards Linde showed that the inflaton
potential need not have either a local minimum or a gentle
plateau: in the scenario he dubbed {\it chaotic inflation}
\cite{chaotic}, the inflaton potential can be as simple as
\begin{equation}
   V(\phi)={1 \over 2} m^2 \phi^2, 
   \label{eq:1}
\end{equation}
provided that $\phi$ begins at a large enough value so that
inflation can occur as it relaxes.  For simplicity of language, I
will stretch the meaning of the phrase ``false vacuum'' to
include all of these cases; that is, I will use the phrase to
denote any state with a high energy density that cannot be
rapidly decreased.  Note that while inflation was originally
developed in the context of grand unified theories, the only real
requirement on the particle physics is the existence of a false
vacuum state.

\subsection{The New Inflationary Scenario:}

In this section I will summarize the workings of new inflation,
and in the following section I will discuss chaotic inflation. 
While more complicated possibilities (e.g. hybrid inflation
\cite{hyb1,hyb2,hyb3,hyb4,hyb5} and supernatural inflation
\cite{RSG}) appear very plausible, the basic scenarios of new and
chaotic inflation will be sufficient to illustrate the physical
effects that I want to discuss in this article.

Suppose that the energy density of a state is approximately equal
to a constant value $\rho_f$.  Then, if a region filled with this
state of matter expanded by an amount $dV$, its energy would have
to increase by
\begin{equation}
  d U = \rho_f \, d V \ . 
  \label{eq:2}
\end{equation}
This energy must be supplied by whatever force is causing the
expansion, which means that the force must be pulling against a
negative pressure.  The work done by the force is given by
\begin{equation}
  dW = - p_f \, d V \ ,
  \label{eq:3}
\end{equation}
where $p_f$ is the pressure inside the expanding region. 
Equating the work with the change in energy, one finds
\begin{equation}
  p_f = - \rho_f \ . 
  \label{eq:4}
\end{equation}
This negative pressure is the driving force behind inflation. 
When one puts this negative pressure into Einstein's equations,
one finds that it leads to a repulsion, causing such a region to
undergo exponential expansion.  If the region can be approximated
as isotropic and homogeneous, this result can be seen from the
standard Friedmann-Robertson-Walker (FRW) equations:
\begin{equation}
  {d^2 a \over d t^2} = - {4 \pi \over 3} G ( \rho + 3 p ) a \ 
   = { 8 \pi \over 3 } G \rho_f a  \ .
  \label{eq:5}
\end{equation}
where $a(t)$ is the scale factor, $G$ is Newton's constant, and
we adopt units for which $\hbar = c = 1$.  For late times the
growing solution to this equation has the form
\begin{equation}
  a(t) \propto e^{\chi t} \ , \hbox{ where } \chi = \sqrt{{8 \pi
   \over 3} G \rho_f } \ .
  \label{eq:6}
\end{equation}
Of course inflationary theorists prefer not to assume that the
universe began homogeneously and isotropically, but there is
considerable evidence for the ``cosmological no-hair conjecture''
\cite{Jensen-Stein-Schabes}, which implies that a wide class of
initial states will approach this exponentially expanding
solution. 

The basic scenario of new inflation begins by assuming that at
least some patch of the early universe was in this peculiar false
vacuum state.  In the original papers
\cite{Linde1,Albrecht-Steinhardt1} this initial
condition was motivated by the fact that, in many quantum field
theories, the false vacuum resulted naturally from the
supercooling of an initially hot state in thermal equilibrium. 
It was soon found, however, that quantum fluctuations in the
rolling inflaton field give rise to density perturbations in the
universe \cite{Starobinsky2,GuthPi,Hawking1,BST,BFM}, and that
these density perturbations would be much larger than observed
unless the inflaton field is very weakly coupled.  For such weak
coupling there would be no time for an initially nonthermal state
to reach thermal equilibrium.  Nonetheless, since thermal
equilibrium describes a probability distribution in which all
states of a given energy are weighted equally, the fact that
thermal equilibrium leads to a false vacuum implies that there
are many ways of reaching a false vacuum.  Thus, even in the
absence of thermal equilibrium---even if the universe started in
a highly chaotic initial state---it seems reasonable to assume
that some small patches of the early universe settled into the
false vacuum state, as was suggested for example in
Ref.~\cite{Guth-RS}.  Linde \cite{chaotic} pointed out that even
highly improbable initial patches could be important if they
inflated, since the exponential expansion could still cause such
patches to dominate the volume of the universe.  One might hope
ultimately to calculate the probability of regions settling into
the false vacuum from a quantum description of cosmogenesis, but
I will argue in Sec.~\ref{implications} that this probability is
quite irrelevant in the context of eternal inflation.

Once a region of false vacuum materializes, the physics of the
subsequent evolution is rather straightforward. The gravitational
repulsion caused by the negative pressure will drive the region
into a period of exponential expansion.  If the energy density of
the false vacuum is at the grand unified theory scale ($\rho_f
\approx (2 \times 10^{16}\ \hbox{GeV})^4)$, Eq.~(\ref{eq:6})
shows that the time constant $\chi^{-1}$ of the exponential
expansion would be about $10^{-38}$ sec. For inflation to achieve
its goals, this patch has to expand exponentially for at least 60
e-foldings.  Then, because the false vacuum is only metastable
(the inflaton field is perched on top of the hill of the
potential energy diagram of Fig.~\ref{newinf}), eventually it
will decay.  The inflaton field will roll off the hill, ending
inflation.  When it does, the energy density that has been locked
in the inflaton field is released. Because of the coupling of the
inflaton to other fields, that energy becomes thermalized to
produce a hot soup of particles, which is exactly what had always
been taken as the starting point of the standard big bang theory
before inflation was introduced.  From here on the scenario joins
the standard big bang description.  The role of inflation is to
establish dynamically the initial conditions which otherwise have
to be postulated. 

The inflationary mechanism produces an entire universe starting
from essentially nothing, so one needs to answer the question of
where the energy of the universe comes from.  The answer is that
it comes from the gravitational field.  The universe did not
begin with this colossal energy stored in the gravitational
field, but rather the gravitational field can supply the energy
because its energy can become negative without bound.  As more
and more positive energy materializes in the form of an
ever-growing region filled with a high-energy scalar field, more
and more negative energy materializes in the form of an expanding
region filled with a gravitational field.  The total energy
remains constant at some very small value, and could in fact be
exactly zero.  There is nothing known that places any limit on
the amount of inflation that can occur while the total energy
remains exactly zero.\footnote{In Newtonian mechanics the energy
density of a gravitational field is unambiguously negative; it
can be derived by the same methods used for the Coulomb field,
but the force law has the opposite sign.  In general relativity
there is no coordinate-invariant way of expressing the energy in
a space that is not asymptotically flat, so many experts prefer
to say that the total energy is undefined.  Either way, there is
agreement that inflation is consistent with the general
relativistic description of energy conservation.}

\subsection{Chaotic Inflation:}

\begin{figure}
\epsfxsize=275pt
\centerline{\epsfbox{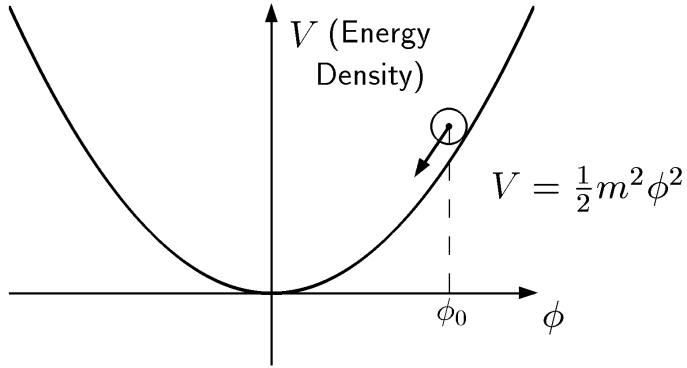}}
\caption{Generic form of the potential for the chaotic inflationary
scenario.}
\label{chaoticinf}
\end{figure}

Chaotic inflation \cite{chaotic} can occur in the context of a
more general class of potential energy functions.  In particular,
even a potential energy function as simple as
Eq.~(\ref{eq:1})---describing a scalar field with a mass and no
interaction---is sufficient to describe chaotic inflation. 
Chaotic inflation is illustrated in Fig.~\ref{chaoticinf}.  In
this case there is no state that bears any obvious resemblance to
the false vacuum of new inflation.  Instead the scenario works by
supposing that chaotic conditions in the early universe produced
one or more patches in which the inflaton field $\phi$ was at
some high value $\phi = \phi_0$ on the potential energy curve. 
Inflation occurs as the inflaton field rolls down the hill.  As
long as the initial value $\phi_0$ is sufficiently large, there
will be sufficient inflation to solve all the problems that
inflation is intended to solve. 

The equations describing chaotic inflation can be written simply,
provided that we assume that the universe is already flat enough
so that we do not need to include a curvature term.  The field
equation for the inflaton field in the expanding universe is
\begin{equation}
  \ddot \phi + 3 H \dot \phi = - {\partial V \over \partial
          \phi } \ ,
  \label{eq:7}
\end{equation}
where the overdot denotes a derivative with respect to time $t$,
and $H$ is the time-dependent Hubble parameter given by
\begin{equation}
  H^2 = {8 \pi \over 3} G V  \ .
  \label{eq:8}
\end{equation}
For the toy-model potential energy of Eq.~(\ref{eq:1}), these
equations have a very simple solution:
\begin{equation}
  \phi = \phi_0 - {m \over \sqrt{12 \pi G}} \, t \ .
  \label{eq:9}
\end{equation}
One can then calculate the number $N$ of inflationary e-foldings,
which is given by
\begin{equation}
  N = \int\nolimits_{\phi = \phi_0}^{\phi = 0} H(t) \, dt = 2 \pi G
     \phi_0^2 \ .
  \label{eq:10}
\end{equation}
In this toy model $N$ depends only on $\phi_0$ and not on the
inflaton mass $m$.  Thus the number of e-foldings will exceed 60
provided that
\begin{equation}
  \phi_0 > \sqrt{60 \over 2 \pi} \, M_{\rm P} \approx 3.1 M_{\rm P} \ ,
  \label{eq:11}
\end{equation}
where $M_{\rm P} \equiv 1/\sqrt{G} = 1.22 \times 10^{19}$ GeV is
the Planck mass.  Although this is a super-Planckian value for
the scalar field, the energy density need not be super-Planckian:
\begin{equation}
  \rho_0 = {1 \over 2} m^2 \phi_0^2 > {60 \over 4 \pi} M_{\rm P}^2 m^2
     \ .
  \label{eq:12}
\end{equation}
For example, if $m = 10^{16}$ GeV, then the potential energy
density is only $3 \times 10^{-6}\, M_{\rm P}^4$.  Since it is
presumably the energy density and not the value of the field that
is relevant to gravity, it seems reasonable to assume that the
chaotic inflation scenario will not be dramatically affected by
corrections from quantum gravity.

\section{Evidence for Inflation}

No matter which form of inflation we might envision, we would
like to know what is the evidence that our universe underwent a
period of inflation.  The answer is pretty much the same no
matter which form of inflation we are discussing.  In my opinion,
the evidence that our universe is the result of some form of
inflation is very solid.  Since the term {\it inflation}
encompasses a wide range of detailed theories, it is hard to
imagine any reasonable alternative.  The basic arguments are as
follows:

\begin{enumerate}
\item{\it The universe is big}

First of all, we know that the universe is incredibly large: the
visible part of the universe contains about $10^{90}$ particles. 
Since we have all grown up in a large universe, it is easy to
take this fact for granted: of course the universe is big, it's
the whole universe! In ``standard'' FRW cosmology, without
inflation, one simply postulates that about $10^{90}$ or more
particles were here from the start.  However, in the context of
present-day cosmology, many of us hope that even the creation of
the universe can be described in scientific terms.  Thus, we are
led to at least think about a theory that might explain how the
universe got to be so big.  Whatever that theory is, it has to
somehow explain the number of particles, $10^{90}$ or more. 
However, it is hard to imagine such a number arising from a
calculation in which the input consists only of geometrical
quantities, quantities associated with simple dynamics, and
factors of 2 or $\pi$.  The easiest way by far to get a huge
number, with only modest numbers as input, is for the calculation
to involve an exponential.  The exponential expansion of
inflation reduces the problem of explaining $10^{90}$ particles
to the problem of explaining 60 or 70 e-foldings of inflation. 
In fact, it is easy to construct underlying particle theories
that will give far more than 70 e-foldings of inflation. 
Inflationary cosmology therefore suggests that, even though the
observed universe is incredibly large, it is only an
infinitesimal fraction of the entire universe.

\item{\it The Hubble expansion}

The Hubble expansion is also easy to take for granted, since we
have all known about it from our earliest readings in cosmology. 
In standard FRW cosmology, the Hubble expansion is part of the
list of postulates that define the initial conditions.  But
inflation actually offers the possibility of explaining how the
Hubble expansion began.  The repulsive gravity associated with
the false vacuum is just what Hubble ordered.  It is exactly the
kind of force needed to propel the universe into a pattern of
motion in which each pair of particles is moving apart with a
velocity proportional to their separation.

\item{\it Homogeneity and isotropy}

The degree of uniformity in the universe is startling.  The
intensity of the cosmic background radiation is the same in all
directions, after it is corrected for the motion of the Earth, to
the incredible precision of one part in 100,000.  To get some
feeling for how high this precision is, we can imagine a marble
that is spherical to one part in 100,000.  The surface of the
marble would have to be shaped to an accuracy of about 1,000
angstroms, a quarter of the wavelength of light. 

Although modern technology makes it possible to grind lenses to
quarter-wavelength accuracy, we would nonetheless be shocked if
we unearthed a stone, produced by natural processes, that was
round to an accuracy of 1,000 angstroms.  If we try to imagine
that such a stone were found, I am sure that no one would accept
an explanation of its origin which simply proposed that the stone
started out perfectly round.  Similarly, I do not think it makes
sense to consider any theory of cosmogenesis that cannot offer
some explanation of how the universe became so incredibly
isotropic. 

The cosmic background radiation was released about 300,000 years
after the big bang, after the universe cooled enough so that the
opaque plasma neutralized into a transparent gas.  The cosmic
background radiation photons have mostly been traveling on
straight lines since then, so they provide an image of what the
universe looked like at 300,000 years after the big bang.  The
observed uniformity of the radiation therefore implies that the
observed universe had become uniform in temperature by that time. 
In standard FRW cosmology, a simple calculation shows that the
uniformity could be established so quickly only if signals could
propagate at 100 times the speed of light, a proposition clearly
contradicting the known laws of physics.  In inflationary
cosmology, however, the uniformity is easily explained.  The
uniformity is created initially on microscopic scales, by normal
thermal-equilibrium processes, and then inflation takes over and
stretches the regions of uniformity to become large enough to
encompass the observed universe.

\item{\it The flatness problem}

I find the flatness problem particularly impressive, because of
the extraordinary numbers that it involves. The problem concerns
the value of the ratio
\begin{equation}
  \Omega_\tot \equiv {\rho_\tot \over \rho_c} \ ,
  \label{eq:13}
\end{equation}
where $\rho_\tot$ is the average total mass density of the
universe and $\rho_c = 3 H^2 / 8 \pi G$ is the critical density,
the density that would make the universe spatially flat.  (In the
definition of ``total mass density,'' I am including the vacuum
energy $\rho_{\rm vac} = \Lambda/ 8 \pi G$ associated with the
cosmological constant $\Lambda$, if it is nonzero.)

There is general agreement that the present value of
$\Omega_\tot$ satisfies
\begin{equation}
  0.1 \lta \Omega_0 \lta 2 \ ,
  \label{eq:14}
\end{equation}
but it is hard to pinpoint the value with more precision. Despite
the breadth of this range, the value of $\Omega$ at early times
is highly constrained, since $\Omega=1$ is an unstable
equilibrium point of the standard model evolution.  Thus, if
$\Omega$ was ever {\it exactly} equal to one, it would remain
exactly one forever.  However, if $\Omega$ differed slightly from
one in the early universe, that difference---whether positive or
negative---would be amplified with time.  In particular, it can
be shown that $\Omega - 1$ grows as
\begin{equation}
  \Omega - 1 \propto \cases{t &(during the radiation-dominated era)\cr
    t^{2/3} &(during the matter-dominated era)\ .\cr}
  \label{eq:15}
\end{equation}
At $t=1$ sec, for example, when the processes of big bang
nucleosynthesis were just beginning, Dicke and Peebles
\cite{dicke} pointed out that $\Omega$ must have equaled one to
an accuracy of one part in $10^{15}$.  Classical cosmology
provides no explanation for this fact---it is simply assumed as
part of the initial conditions.  In the context of modern
particle theory, where we try to push things all the way back to
the Planck time, $10^{-43}$ sec, the problem becomes even more
extreme.  If one specifies the value of $\Omega$ at the Planck
time, it has to equal one to 58 decimal places in order to be
anywhere in the allowed range today. 

While this extraordinary flatness of the early universe has no
explanation in classical FRW cosmology, it is a natural
prediction for inflationary cosmology.  During the inflationary
period, instead of $\Omega$ being driven away from one as
described by Eq.~(\ref{eq:15}), $\Omega$ is driven towards one,
with exponential swiftness:
\begin{equation}
  \Omega - 1 \propto e^{-2 H_{\rm inf} t} \ ,
  \label{eq:16}
\end{equation}
where $H_{\rm inf}$ is the Hubble parameter during inflation. 
Thus, as long as there is a long enough period of inflation,
$\Omega$ can start at almost any value, and it will be driven to
unity by the exponential expansion. 

\item{\it Absence of magnetic monopoles}

All grand unified theories predict that there should be, in the
spectrum of possible particles, extremely massive particles
carrying a net magnetic charge.  By combining grand unified
theories with classical cosmology without inflation, Preskill
\cite{preskill} found that magnetic monopoles would be produced
so copiously that they would outweigh everything else in the
universe by a factor of about $10^{12}$.  A mass density this
large would cause the inferred age of the universe to drop to
about 30,000 years! Inflation is certainly the simplest known
mechanism to eliminate monopoles from the visible universe, even
though they are still in the spectrum of possible particles.  The
monopoles are eliminated simply by arranging the parameters so
that inflation takes place after (or during) monopole production,
so the monopole density is diluted to a completely negligible
level.

\item{\it Anisotropy of the cosmic background radiation}

The process of inflation smooths the universe essentially
completely, but density fluctuations are generated as inflation
ends by the quantum fluctuations of the inflaton field. 
Generically these are adiabatic Gaussian fluctuations with a
nearly scale-invariant spectrum
\cite{Starobinsky2,GuthPi,Hawking1,BST,BFM}.  New data is
arriving quickly, but so far the observations are in excellent
agreement with the predictions of the simplest inflationary
models.  For a review, see for example Bond and Jaffe
\cite{bond-jaffe}, who find that the combined data give a slope
of the primordial power spectrum within 5\% of the preferred
scale-invariant value.

\end{enumerate}

\section{Eternal Inflation: Mechanisms}

The remainder of this article will discuss eternal
inflation---the questions that it can answer, and the questions
that it raises.  In this section I discuss the mechanisms that
make eternal inflation possible, leaving the other issues for the
following sections.  I will discuss eternal inflation first in
the context of new inflation, and then in the context of chaotic
inflation, where it is more subtle. 

\subsection{Eternal New Inflation:}

The eternal nature of new inflation was first discovered by
Steinhardt \cite{steinhardt-nuffield} and Vilenkin
\cite{vilenkin-eternal} in 1983.  Although the false vacuum is a
metastable state, the decay of the false vacuum is an exponential
process, very much like the decay of any radioactive or unstable
substance.  The probability of finding the inflaton field at the
top of the plateau in its potential energy diagram does not fall
sharply to zero, but instead trails off exponentially with time
\cite{guth-pi2}.  However, unlike a normal radioactive substance,
the false vacuum exponentially expands at the same time that it
decays. In fact, in any successful inflationary model the rate of
exponential expansion is always much faster than the rate of
exponential decay.  Therefore, even though the false vacuum is
decaying, it never disappears, and in fact the total volume of
the false vacuum, once inflation starts, continues to grow
exponentially with time, ad infinitum. 

\begin{figure}
\centerline{\epsfbox{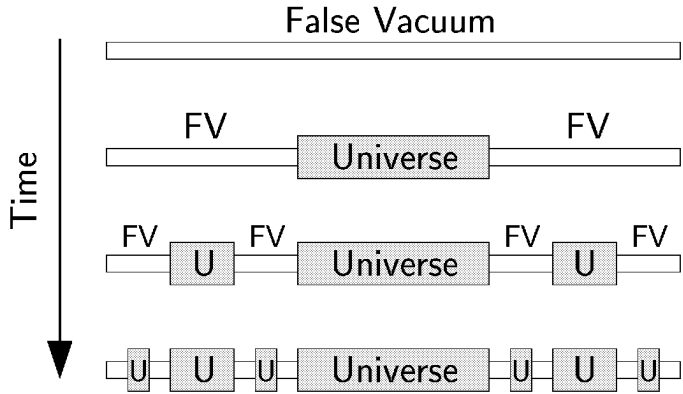}}
\caption{A schematic illustration of eternal inflation.} 
\label{eternalline}
\end{figure}

Fig.~\ref{eternalline} shows a schematic diagram of an eternally
inflating universe.  The top bar indicates a region of false
vacuum.  The evolution of this region is shown by the successive
bars moving downward, except that the expansion could not be
shown while still fitting all the bars on the page.  So the
region is shown as having a fixed size in comoving coordinates,
while the scale factor, which is not shown, increases from each
bar to the next.  As a concrete example, suppose that the scale
factor for each bar is three times larger than for the previous
bar.  If we follow the region of false vacuum as it evolves from
the situation shown in the top bar to the situation shown in the
second bar, in about one third of the region the scalar field
rolls down the hill of the potential energy diagram,
precipitating a local big bang that will evolve into something
that will eventually appear to its inhabitants as a universe. 
This local big bang region is shown in gray and labelled
``Universe.'' Meanwhile, however, the space has expanded so much
that each of the two remaining regions of false vacuum is the
same size as the starting region.  Thus, if we follow the region
for another time interval of the same duration, each of these
regions of false vacuum will break up, with about one third of
each evolving into a local universe, as shown on the third bar
from the top.  Now there are four remaining regions of false
vacuum, and again each is as large as the starting region.  This
process will repeat itself literally forever, producing a kind of
a fractal structure to the universe, resulting in an infinite
number of the local universes shown in gray.  There is no
standard name for these local universes, but they are often
called bubble universes.  I prefer, however, to call them pocket
universes, to avoid the suggestion that they are round.  While
bubbles formed in first-order phase transitions are round
\cite{coleman-deluccia}, the local universes formed in eternal
new inflation are generally very irregular, as can be seen for
example in the two-dimensional simulation by Vanchurin, Vilenkin,
and Winitzki in Fig.~2 of Ref.~\cite{vvw}.

The diagram in Fig.~\ref{eternalline} is of course an
idealization.  The real universe is three dimensional, while the
diagram illustrates a schematic one-dimensional universe.  It is
also important that the decay of the false vacuum is really a
random process, while the diagram was constructed to show a very
systematic decay, because it is easier to draw and to think
about.  When these inaccuracies are corrected, we are still left
with a scenario in which inflation leads asymptotically to a
fractal structure \cite{aryal-vilenkin} in which the universe as
a whole is populated by pocket universes on arbitrarily small
comoving scales.  Of course this fractal structure is entirely on
distance scales much too large to be observed, so we cannot
expect astronomers to see it.  Nonetheless, one does have to
think about the fractal structure if one wants to understand the
very large scale structure of the spacetime produced by
inflation. 

Most important of all is the simple statement that once inflation
happens, it produces not just one universe, but an infinite
number of universes. 

\subsection{Eternal Chaotic Inflation:}

The eternal nature of new inflation depends crucially on the
scalar field lingering at the top of the plateau of
Fig.~\ref{newinf}.  Since the potential function for chaotic
inflation, Fig.~\ref{chaoticinf}, does not have a plateau, it is
not obvious how eternal inflation can happen in this context. 
Nonetheless, Andrei Linde \cite{linde-eternal} showed in 1986
that chaotic inflation can also be eternal. 

The important point is that quantum fluctuations play an
important role in all inflationary models.  Quantum fluctuations
are invariably important on very small scales, and with inflation
these very small scales are rapidly stretched to become
macroscopic and even astronomical.  Thus the scalar field
associated with inflation has very evident quantum effects. 

\begin{figure}
\epsfxsize=275pt
\centerline{\epsfbox{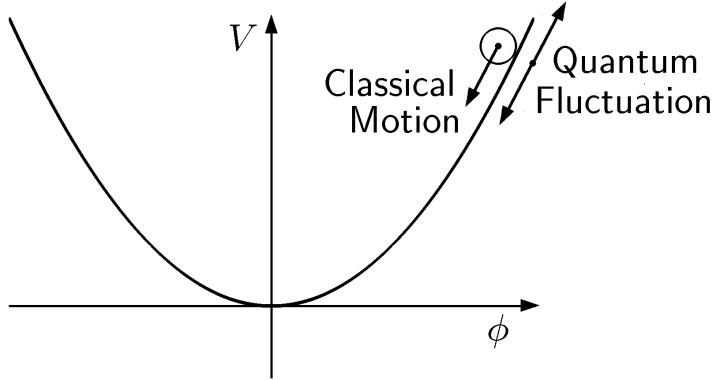}}
\caption{Evolution of the inflaton field during eternal chaotic
inflation.}
\label{chaotic-eternal}
\end{figure}

When the mass of the scalar field is small compared to the Hubble
parameter $H$, these quantum effects are accurately summarized by
saying that the quantum fluctuations cause the field to undergo a
random walk.  It is useful to divide space into regions of
physical size $H^{-1}$, and to discuss the average value of the
scalar field $\phi$ within a given region.  In a time $H^{-1}$,
the effect of the quantum fluctuations is equivalent to a random
Gaussian jump of zero mean and a root-mean-squared magnitude
\cite{random-vil-ford,random-linde,Starobinsky2,random-starobinsky}
given by
\begin{equation}
  \Delta \phi_{\rm qu} = {H \over 2 \pi} \ .
  \label{eq:17}
\end{equation}
This random quantum jump is superimposed on the classical motion,
as indicated in Fig.~(\ref{chaotic-eternal}).

To illustrate how eternal inflation happens in the simplest
context, let us consider again the free scalar field described by
the potential function of Eq.~(\ref{eq:1}).  We consider a region
of physical radius $H^{-1}$, in which the field has an average
value $\phi$.  Using Eq.~(\ref{eq:9}) along with
Eqs.~(\ref{eq:8}) and (\ref{eq:1}), one finds that the magnitude
of the classical change that the field will undergo in a time
$H^{-1}$ is given in by
\begin{equation}
  \Delta \phi_{\rm cl} = {M_{\rm P} m \over \sqrt{12 \pi}} \,
     H^{-1} = {1 \over 4 \pi} {M_{\rm P}^2 \over \phi} \ .
  \label{eq:18}
\end{equation}
Let $\phi^*$ denote the value of $\phi$ which is sufficiently
large so that
\begin{equation}
  \Delta \phi_{\rm qu} (\phi^*) = \Delta \phi_{\rm cl}(\phi^*) \ ,
  \label{eq:19}
\end{equation}
which can easily be solved to find
\begin{equation}
  \phi^* = \left( {3 \over 16 \pi} \right)^{1/4} {M_{\rm P}^{3/2} \over
     m^{1/2}} \ .
  \label{eq:20}
\end{equation}

Now consider what happens to the region if its initial average
value of $\phi$ is equal to $\phi^*$.  In a time interval
$H^{-1}$, the volume of the region will increase by $e^3 \approx
20$.  At the end of the time interval we can divide the original
region into 20 regions of the same volume as the original, and in
each region the average scalar field can be written as
\begin{equation}
  \phi = \phi^* + \Delta \phi_{\rm cl} + \delta \phi \ ,
  \label{eq:21}
\end{equation}
where $\delta \phi$ denotes the random quantum jump, which is
drawn from a Gaussian probability distribution with standard
deviation $\Delta \phi_{\rm qu} = \Delta \phi_{\rm cl}$. 
Gaussian statistics imply that there is a 15.9\% chance that a
Gaussian random variable will exceed its mean by more than one
standard deviation, and therefore there is a 15.9\% chance that
the net change in $\phi$ will be positive.  Since there are now
20 regions of the original volume, on average the value of $\phi$
will exceed the original value in 3.2 of these regions.  Thus the
volume for which $\phi \ge \phi^*$ does not (on average)
decrease, but instead increases by more than a factor of 3. 
Since this argument can be iterated, the expectation value of the
volume for which $\phi \ge \phi^*$ increases exponentially with
time.  Typically, therefore, inflation never ends, but instead
the volume of the inflating region grows exponentially without
bound.  The minimum field value for eternal inflation is slightly
below $\phi^*$, since a volume increase by a factor of 3.2 is
more than necessary---any factor greater than one would be
sufficient.  A short calculation shows that the minimal value for
eternal inflation is $0.78 \phi^*$.

While the value of $\phi^*$ is larger than $M_{\rm P}$, it is
important to note that the energy density can still be much
smaller than Planck scale:
\begin{equation}
  V(\phi^*) = {1 \over 2} m^2 \phi^{*2} = \sqrt{3 \over 64 \pi}
     \, m M_{\rm P}^3 \ ,
  \label{eq:22}
\end{equation}
which for $m = 10^{16}$ GeV gives an energy density of $1 \times
10^{-4} \, M_{\rm P}^4$. 

If one repeats the argument with a potential function
\begin{equation}
  V(\phi) = {1 \over 4} \lambda \phi^4 \ ,
  \label{eq:23}
\end{equation}
one finds \cite{linde-book} that
\begin{equation}
  \phi^* = \left( 3 \over 2 \pi \lambda \right)^{1/6} M_{\rm P} \ ,
  \label{eq:24}
\end{equation}
and
\begin{equation}
  V(\phi^*) = \left( 3 \over 16 \pi \right)^{2/3} \lambda^{1/3}
     M_{\rm P}^4 \ .
  \label{eq:25}
\end{equation}
Since one requires $\lambda$ to be very small in any case so that
density perturbations are not too large, one finds again that
eternal inflation is predicted to happen at an energy density
well below the Planck scale.

\section{Eternal Inflation: Implications}
\label{implications}

In spite of the fact that the other universes created by eternal
inflation are too remote to imagine observing directly, I
nonetheless claim that eternal inflation has real consequences in
terms of the way we extract predictions from theoretical models. 
Specifically, there are three consequences of eternal inflation
that I will discuss.

First, eternal inflation implies that all hypotheses about the
ultimate initial conditions for the universe---such as the
Hartle-Hawking \cite{hartle-hawking} no boundary proposal, the
tunneling proposals by Vilenkin \cite{tunnel-vilenkin} or Linde
\cite{tunnel-linde}, or the more recent Hawking-Turok instanton
\cite{hawking-turok}---become totally divorced from observation.
That is, one would expect that if inflation is to continue
arbitrarily far into the future with the production of an
infinite number of pocket universes, then the statistical
properties of the inflating region should approach a steady state
which is independent of the initial conditions.  Unfortunately,
attempts to quantitatively study this steady state are severely
limited by several factors.  First, there are ambiguities in
defining probabilities, which will be discussed later.  In
addition, the steady state properties seem to depend strongly on
super-Planckian physics which we do not understand.  That is, the
same quantum fluctuations that make eternal chaotic inflation
possible tend to drive the scalar field further and further up
the potential energy curve, so attempts to quantify the steady
state probability distribution \cite{LLM,GBLinde} require the
imposition of some kind of a boundary condition at large $\phi$. 
Although these problems remain unsolved, I still believe that it
is reasonable to assume that in the course of its unending
evolution, an eternally inflating universe would lose all memory
of the state in which it started.

Even if the universe forgets the details of its genesis, however,
I would not assume that the question of how the universe began
would lose its interest.  While eternally inflating universes
continue forever once they start, they are presumably not eternal
into the past.  (The word {\it eternal} is therefore not
technically correct---it would be more precise to call this
scenario {\it semi-eternal} or {\it future-eternal}.) While the
issue is not completely settled, it appears likely that eternally
inflating universes must necessarily have a beginning.  Borde and
Vilenkin \cite{borde-vilenkin} have shown, subject to various
assumptions, that spacetimes that are future-eternal must have an
initial singularity, in the sense that they cannot be past null
geodesically complete.  The proof, however, requires the weak
energy condition, which can be violated by quantum fluctuations
\cite{borde-vilenkin2}.  In any case, no one has constructed a
viable model without a beginning, and certainly nothing that we
know can rule out the possibility of a beginning.  The
possibility of a quantum origin of the universe is very
attractive, and will no doubt be a subject of interest for some
time.  Eternal inflation, however, seems to imply that the entire
study will have to be conducted with literally no input from
observation. 

A second consequence of eternal inflation is that the probability
of the onset of inflation becomes totally irrelevant, provided
that the probability is not identically zero. Various authors in
the past have argued that one type of inflation is more plausible
than another, because the initial conditions that it requires
appear more likely to have occurred.  In the context of eternal
inflation, however, such arguments have no significance.

To illustrate the insignificance of the probability of the onset
of inflation, I will use a numerical example.  We will imagine
comparing two different versions of inflation, which I will call
Type A and Type B\relax.  They are both eternally inflating---but
Type A will have a higher probability of starting, while Type B
will be a little faster in its exponential expansion rate.  Since
I am trying to show that the higher starting probability of Type
A is irrelevant, I will choose my numbers to be extremely
generous to Type A\relax.  First, we must choose a number for how
much more probable it is for Type A inflation to begin, relative
to type B\relax.  A googol, $10^{100}$, is usually considered a
large number---it is some 20 orders of magnitude larger than the
total number of baryons in the visible universe.  But I will be
more generous: I will assume that Type A inflation is more likely
to start than type B inflation by a factor of $10^{1,000,000}$. 
Type B inflation, however, expands just a little bit faster, say
by 0.001\%.  We need to choose a time constant for the
exponential expansion, which I will take to be a typical grand
unified theory scale, $\tau = 10^{-37}$ sec. ($\tau$ represents
the time constant for the overall expansion factor, which takes
into account both the inflationary expansion and the exponential
decay of the false vacuum.) Finally, we need to choose a length
of time to let the system evolve.  In principle this time
interval is infinite (the inflation is eternal into the future),
but to be conservative we will follow the system for only one
second.

We imagine starting a statistical ensemble of universes at $t=0$,
with an expectation value for the volume of Type A inflation
exceeding that of Type B inflation by $10^{1,000,000}$.  For
brevity, I will use the term ``weight'' to refer to the ensemble
expectation value of the volume.  Thus, at $t=0$ the weights of
Type A inflation and Type B inflation will have the ratio
\begin{equation}
  \left.{W_B \over W_A}\right|_{t=0} = 10^{-1,000,000} \ .
  \label{eq:26}
\end{equation}
After one second of evolution, the expansion factors for Type A
and Type B inflation will be
\begin{eqnarray}
  \label{eq:27}
  Z_A & = & e^{t/\tau} = e^{10^{37}} \\
  \label{eq:28}
  Z_B & = & e^{1.00001\,t/\tau} = e^{0.00001\,t/\tau} Z_A
          \nonumber \\
      & = & e^{10^{32}} Z_A \approx 10^{4.3 \times 10^{31}} Z_A
\end{eqnarray}
The weights at the end of one second are proportional to these
expansion factors, so
\begin{equation}
  \left.{W_B \over W_A}\right|_{t=1\ \rm sec} = 10^{\left(4.3
     \times 10^{31} - 1,000,000\right)} \ .
  \label{eq:29}
\end{equation}
Thus, the initial ratio of $10^{1,000,000}$ is vastly superseded
by the difference in exponential expansion factors.  In fact, we
would have to calculate the exponent of Eq.~(\ref{eq:29}) to an
accuracy of 25 significant figures to be able to barely detect
the effect of the initial factor of $10^{1,000,000}$. 

One might criticize the above argument for being naive, as the
concept of time was invoked without any specification of how the
equal-time hypersurfaces are to be defined.  I do not know a
decisive answer to this objection; as I will discuss later, there
are unresolved questions concerning the calculation of
probabilities in eternally inflating spacetimes.  Nonetheless,
given that there is actually an infinity of time available, it is
seems reasonable to believe that the form of inflation that
expands the fastest will always dominate over the slower forms by
an infinite factor.

A corollary to this argument is that new inflation is not dead. 
While the initial conditions necessary for new inflation cannot
be justified on the basis of thermal equilibrium, as proposed in
the original papers \cite{Linde1,Albrecht-Steinhardt1}, in the
context of eternal inflation it is sufficient to conclude that
the probability for the required initial conditions is nonzero. 
Since the resulting scenario does not depend on the words that
are used to justify the initial state, the standard treatment of
new inflation remains valid.

A third consequence of eternal inflation is the possibility that
it offers to rescue the predictive power of theoretical physics. 
Here I have in mind the status of string theory, or the theory
known as M theory, into which string theory has evolved.  The
theory itself has an elegant uniqueness, but nonetheless it is
not at all clear that the theory possesses a unique vacuum. 
Since predictions will ultimately depend on the properties of the
vacuum, the predictive power of string/M theory may be limited. 
Eternal inflation, however, provides a possible mechanism to
remedy this problem.  Even if many types of vacua are equally
stable, it may turn out that one of them leads to a maximal rate
of inflation.  If so, then this type of vacuum will dominate the
universe, even if its expansion rate is only infinitesimally
larger than the other possibilities.  Thus, eternal inflation
might allow physicists to extract unique predictions, in spite of
the multiplicity of stable vacua.

\section{Difficulties in Calculating Probabilities}

In an eternally inflating universe, anything that can happen will
happen; in fact, it will happen an infinite number of times. 
Thus, the question of what is possible becomes trivial---anything
is possible, unless it violates some absolute conservation law. 
To extract predictions from the theory, we must therefore learn
to distinguish the probable from the improbable.

However, as soon as one attempts to define probabilities in an
eternally inflating spacetime, one discovers ambiguities.  The
problem is that the sample space is infinite, in that an
eternally inflating universe produces an infinite number of
pocket universes.  The fraction of universes with any particular
property is therefore equal to infinity divided by infinity---a
meaningless ratio.  To obtain a well-defined answer, one needs to
invoke some method of regularization.

To understand the nature of the problem, it is useful to think
about the integers as a model system with an infinite number of
entities.  We can ask, for example, what fraction of the integers
are odd.  Most people would presumably say that the answer is
$1/2$, since the integers alternate between odd and even.  That
is, if the string of integers is truncated after the $N$th, then
the fraction of odd integers in the string is exactly $1/2$ if
$N$ is even, and is $(N+1)/2N$ if $N$ is odd.  In any case, the
fraction approaches $1/2$ as $N$ approaches infinity.

However, the ambiguity of the answer can be seen if one imagines
other orderings for the integers.  One could, if one wished,
order the integers as
\begin{equation}
  1,3,\ 2,\ 5,7,\ 4,\ 9,11,\ 6\ ,\ldots,  
  \label{eq:30}
\end{equation}
always writing two odd integers followed by one even integer. 
This series includes each integer exactly once, just like the
usual sequence ($1,2,3,4, \ldots$).  The integers are just
arranged in an unusual order.  However, if we truncate the
sequence shown in Eq.~(\ref{eq:30}) after the $N$th entry, and
then take the limit $N \to \infty$, we would conclude that 2/3 of
the integers are odd.  Thus, we find that the definition of
probability on an infinite set requires some method of
truncation, and that the answer can depend nontrivially on the
method that is used.

In the case of eternally inflating spacetimes, the natural choice
of truncation might be to order the pocket universes in the
sequence in which they form.  However, we must remember that each
pocket universe fills its own future light cone, so no pocket
universe forms in the future light cone of another.  Any two
pocket universes are spacelike separated from each other, so some
observers will see one as forming first, while other observers
will see the opposite.  One can arbitrarily choose equal-time
surfaces that foliate the spacetime, and then truncate at some
value of $t$, but this recipe is not unique.  In practice,
different ways of choosing equal-time surfaces give different
results. 

\section{The Youngness Paradox}

If one chooses a truncation in the most naive way, one is led to
a set of very peculiar results which I call the {\it youngness
paradox.}

Specifically, suppose that one constructs a Robertson-Walker
coordinate system while the model universe is still in the false
vacuum (de Sitter) phase, before any pocket universes have
formed. One can then propagate this coordinate system forward
with a synchronous gauge condition,\footnote{By a synchronous
gauge condition, I mean that each equal-time hypersurface is
obtained by propagating every point on the previous hypersurface
by a fixed infinitesimal time interval $\Delta t$ in the
direction normal to the hypersurface.} and one can define
probabilities by truncating at a fixed value $t_f$ of the
synchronous time coordinate $t$.  That is, the probability of any
particular property can be taken to be proportional to the volume
on the $t = t_f$ hypersurface which has that property.  This
method of defining probabilities was studied in detail by Linde,
Linde, and Mezhlumian, in a paper with the memorable title ``Do
we live in the center of the world?'' \cite{center-world}.  I
will refer to probabilities defined in this way as synchronous
gauge probabilities.

The youngness paradox is caused by the fact that the volume of
false vacuum is growing exponentially with time with an
extraordinary time constant, in the vicinity of $10^{-37}$ sec.
Since the rate at which pocket universes form is proportional to
the volume of false vacuum, this rate is increasing exponentially
with the same time constant.  That means that in each second the
number of pocket universes that exist is multiplied by a factor
of $\exp\left\{10^{37}\right\}$.  At any given time, therefore,
almost all of the pocket universes that exist are universes that
formed very very recently, within the last several time
constants.  The population of pocket universes is therefore an
incredibly youth-dominated society, in which the mature universes
are vastly outnumbered by universes that have just barely begun
to evolve.  Although the mature universes have a larger volume,
this multiplicative factor is of little importance, since in
synchronous coordinates the volume no longer grows exponentially
once the pocket universe forms.

Probability calculations in this youth-dominated ensemble lead to
peculiar results, as discussed in Ref.~\cite{center-world}. 
These authors considered the expected behavior of the mass
density in our vicinity, concluding that we should find ourselves
very near the center of a spherical low-density region.  Here I
would like to discuss a less physical but simpler question, just
to illustrate the paradoxes associated with synchronous gauge
probabilities.  Specifically, I will consider the question: ``Are
there any other civilizations in the visible universe that are
more advanced than ours?''.  Intuitively I would not expect
inflation to make any predictions about this question, but I will
argue that the synchronous gauge probability distribution
strongly implies that there is no civilization in the visible
universe more advanced than us.

Suppose that we have reached some level of advancement, and
suppose that $t_{\rm min}$ represents the minimum amount of time
needed for a civilization as advanced as we are to evolve,
starting from the moment of the decay of the false vacuum---the
start of the big bang.  The reader might object on the grounds
that there are many possible measures of advancement, but I would
respond by inviting the reader to pick any measure she chooses;
the argument that I am about to give should apply to all of them. 
The reader might alternatively claim that there is no sharp
minimum $t_{\rm min}$, but instead we should describe the problem
in terms of a function which gives the probability that, for any
given pocket universe, a civilization as advanced as we are would
develop by time $t$. I believe, however, that the introduction of
such a probability distribution would merely complicate the
argument, without changing the result. So, for simplicity of
discussion, I will assume that there is some sharply defined
minimum time $t_{\rm min}$ required for a civilization as
advanced as ours to develop.

Since we exist, our pocket universe must have an age $t_0$
satisfying
\begin{equation}
  t_0 \ge t_{\rm min} \ . 
  \label{eq:31}
\end{equation}
Suppose, however, that there is some civilization in our pocket
universe that is more advanced than we are, let us say by 1
second.  In that case Eq.~(\ref{eq:31}) is not sufficient, but
instead the age of our pocket universe would have to satisfy
\begin{equation}
  t_0 \ge t_{\rm min} + 1 \hbox{\ second}\ . 
  \label{eq:32}
\end{equation}
However, in the synchronous gauge probability distribution,
universes that satisfy Eq.~(\ref{eq:32}) are outnumbered by
universes that satisfy Eq.~(\ref{eq:31}) by a factor of
approximately $\exp\left\{10^{37}\right\}$.  Thus, if we know
only that we are living in a pocket universe that satisfies
Eq.~(\ref{eq:31}), it is extremely improbable that it also
satisfies Eq.~(\ref{eq:32}).  We would conclude, therefore, that
it is extraordinarily improbable that there is a civilization in
our pocket universe that is at least 1 second more advanced than
we are.

Perhaps this argument explains why SETI has not found any signals
from alien civilizations, but I find it more plausible that it is
merely a symptom that the synchronous gauge probability
distribution is not the right one.

\section{An Alternative Probability Prescription}

Since the probability measure depends on the method used to
truncate the infinite spacetime of eternal inflation, we are not
forced to accept the consequences of the synchronous gauge
probabilities.  A very attractive alternative has been proposed
by Vilenkin \cite{vilenkin-proposal}, and developed further by
Vanchurin, Vilenkin, and Winitzki \cite{vvw}.

The key idea of the Vilenkin proposal is to define probabilities
within a single pocket universe (which he describes more
precisely as a connected, thermalized domain).  Thus, unlike the
synchronous gauge method, there is no comparison between old
pocket universes and young ones.  To justify this approach it is
crucial to recognize that each pocket universe is infinite, even
if one starts the model with a finite region of de Sitter space. 
The infinite volume arises in the same way as it does for the
special case of Coleman-de Luccia bubbles
\cite{coleman-deluccia}, the interior of which are open
Robertson-Walker universes.  From the outside one often describes
such bubbles in a coordinate system in which they are finite at
any fixed time, but in which they grow without bound.  On the
inside, however, the natural coordinate system is the one that
reflects the intrinsic homogeneity, in which the space is
infinite at any given time.  The infinity of time, as seen from
the outside, becomes an infinity of spatial extent as seen on the
inside.  Thus, at least for continuously variable parameters, a
single pocket universe provides an infinite sample space which
can be used to define probabilities.  The second key idea of
Vilenkin's method is to use the inflaton field itself as the time
variable, rather than the synchronous time variable discussed in
the previous section.

This approach can be used, for example, to discuss the
probability distribution for $\Omega$ in open inflationary
models, or to discuss the probability distribution for some
arbitrary field that has a flat potential energy function.  If,
however, the vacuum has a discrete parameter which is homogeneous
within each pocket universe, but which takes on different values
in different pocket universes, then this method does not apply. 

\begin{figure}
\centerline{\epsfbox{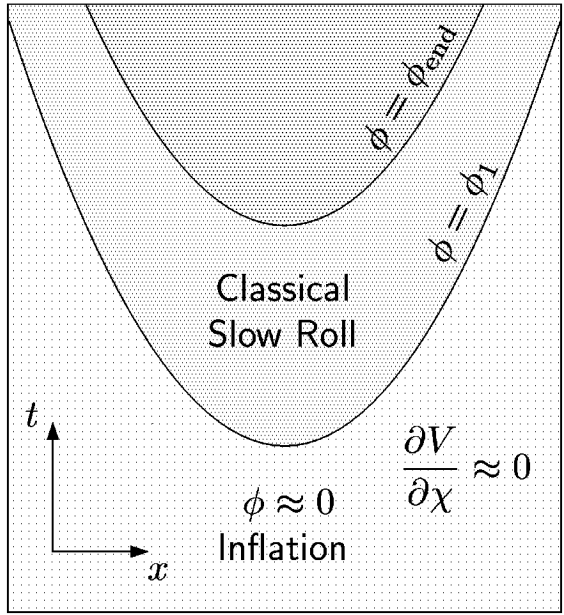}}
\caption{A schematic picture of a pocket universe, illustrating
Vilenkin's proposal for the calculation of probabilities.}
\label{vilenkin-space}
\end{figure}

The proposal can be described in terms of
Fig.~\ref{vilenkin-space}.  We suppose that the theory includes
an inflaton field $\phi$ of the new inflation type, and some set
of fields $\chi_i$ which have flat potentials.  The goal is to
find the probability distribution for the fields $\chi_i$.  We
assume that the evolution of the inflaton $\phi$ can be divided
into three regimes, as shown on the figure.  $\phi < \phi_1$
describes the eternally inflating regime, in which the evolution
is governed by quantum diffusion.  For $\phi_1 < \phi < \phi_{\rm
end}$, the evolution is described classically in a slow-roll
approximation, so that $\dot \phi \equiv {\rm d} \phi / {\rm d}
t$ can be expressed as a function of $\phi$.  For $\phi >
\phi_{\rm end}$ inflation is over, and the $\phi$ field no longer
plays an important role in the evolution.  The $\chi_i$ fields
are assumed to have a finite range of values, such as angular
variables, so that a flat probability distribution is
normalizable.  They are assumed to have a flat potential energy
function for $\phi > \phi_{\rm end}$, so that they could settle
at any value.  They are also assumed to have a flat potential
energy function for $\phi < \phi_1$, although they might interact
with $\phi$ during the slow-roll regime, however, so that they
can affect the rate of inflation. 

Since the potential for the $\chi_i$ is flat for $\phi < \phi_1$,
we can assume that they begin with a flat probability
distribution $P_0(\chi_i) \equiv P(\chi_i, \phi_1)$ on the $\phi
= \phi_1$ hypersurface.  If the kinetic energy function for the
$\chi_i$ is of the standard form, we take $P_0(\chi_i) = const$. 
If, however, the kinetic energy is nonstandard,
\begin{equation}
  {\cal L}_{\rm kinetic} = g^{ij}(\chi) \partial_\mu \chi_i
     \partial^\mu \chi_j \ ,
  \label{eq:33}
\end{equation}
as is plausible for a field described in angular variables, then
the initial probability distribution is assumed to take the
reparameterization-invariant form
\begin{equation}
  P_0(\chi_i) \propto \sqrt{ \det g}  \ .
  \label{eq:34}
\end{equation}
During the slow-roll era, it is assumed that the $\chi_i$ fields
evolve classically, so one can calculate the number of e-folds of
inflation $N(\chi_i)$ as a function of the final value of the
$\chi_i$ (i.e., the value of $\chi_i$ on the $\phi =
\phi_{\rm end}$ hypersurface).  
One can also calculate the final values $\chi_i$ in terms of the
initial values $\chi_i^0$ (i.e., the value of $\chi_i$ on the
$\phi=\phi_1$ hypersurface). One then assumes that the
probability density is enhanced by the volume inflation factor
$e^{3 N (\chi_i)}$, and that the evolution from $\chi_i^0$ to
$\chi_i$ results in a Jacobian factor.  The (unnormalized) final
probability distribution is thus given by
\begin{equation}
  P(\chi_i, \phi_{\rm end}) = P_0(\chi_i^0) e^{3 N (\chi_i)} \,
     \det {\partial \chi_j^0 \over \partial \chi_k} \ .
  \label{eq:35}
\end{equation}
Alternatively, if the evolution of the $\chi_i$ during the
slow-roll era is subject to quantum fluctuations, Ref.~\cite{vvw}
shows how to write a Fokker-Planck equation which is equivalent
to averaging the result of Eq.~(\ref{eq:35}) over a collection of
paths that result from interactions with a noise term.

The Vilenkin proposal sidesteps the youngness paradox by defining
probabilities by the comparison of volumes within one pocket
universe.  The youngness paradox, in contrast, arose when one
considered a probability ensemble of all pocket universes at a
fixed value of the synchronous gauge time coordinate---an
ensemble that is overwhelmingly dominated by very young pocket
universes.

The proposal has the drawback, however, that it cannot be used to
compare the probabilities of discretely different alternatives. 
Furthermore, although the results of this method seem reasonable,
I do not at this point find them compelling.  That is, it is not
clear what principles of physics or probability theory ensure
that this particular method of regularizing the spacetime is the
one that leads to correct predictions.  Perhaps there is no way
to answer this question, so we may be forced to accept this
proposal, or something similar to it, as a postulate.

\section{Conclusion}

In this paper I have summarized the workings of inflation, and
the arguments that strongly suggest that our universe is the
product of inflation.  I argued that inflation can explain the
size, the Hubble expansion, the homogeneity, the isotropy, and
the flatness of our universe, as well as the absence of magnetic
monopoles, and even the characteristics of the nonuniformities. 
The detailed observations of the cosmic background radiation
anisotropies continue to fall in line with inflationary
expectations, and the evidence for an accelerating universe fits
well with the inflationary preference for a flat universe.

Next I turned to the question of eternal inflation, claiming that
essentially all inflationary models are eternal. In my opinion
this makes inflation very robust: if it starts anywhere, at any
time in all of eternity, it produces an infinite number of pocket
universes.  Eternal inflation has the very attractive feature,
from my point of view, that it offers the possibility of allowing
unique predictions even if the underlying string theory does not
have a unique vacuum.  I have also emphasized, however, that
there are important problems in understanding the implications of
eternal inflation.  First, there is the problem that we do not
know how to treat the situation in which the scalar field climbs
upward to the Planck energy scale.  Second, the definition of
probabilities in an eternally inflating spacetime is not yet a
closed issue, although important progress has been made.  And
third, I might add that the entire present approach is at best
semiclassical.  A better treatment may not be possible until we
have a much better handle on quantum gravity, but eventually this
issue will have to be faced.

\section*{Acknowledgments}

The author particularly thanks Andrei Linde, Alexander Vilenkin,
Neil Turok, and other participants in the Isaac Newton Institute
programme {\it Structure Formation in the Universe} for very
helpful conversations.  This work is supported in part by funds
provided by the U.S. Department of Energy (D.O.E.) under
cooperative research agreement \#DF-FC02-94ER40818, and in part
by funds provided by NM Rothschild \& Sons Ltd and by the EPSRC.

% Macros that control the format for all references:
\newcommand{\jf}{\it}  % Journal name font
\newcommand{\jt}{\/}   % Journal name spacing correction
                       % Replace by {} if Journal name font is not italics
\newcommand{\VPY}[3]{{\bf #1}, #2 (#3)}  % Volume Page Year
\newcommand{\ispace}{\thinspace}   % Space between multiple initials

% Macro to implement the \. command for spacing \ispace between initials
% LaTeX uses \.{o} for an o with an overdot, but I want to use \. for
%    periods separating multiple initials.  Use \U{o} for an o with an 
%    overdot:
\let\U=\.
\def\.{.\nobreak\ispace\ignorespaces}

% Macros for individual journal names:
\newcommand{\IJMODPHYS}[3]{{\jf Int. J. Mod. Phys.\jt} \VPY{#1}{#2}{#3}}
\newcommand{\JETP}[3]{{\jf JETP Lett.\jt} \VPY{#1}{#2}{#3}}
\newcommand{\MPL}[3]{{\jf Mod. Phys. Lett.\jt} \VPY{#1}{#2}{#3}}
\newcommand{\NC}[3]{{\jf Nuovo Cim.\jt} \VPY{#1}{#2}{#3}}
\newcommand{\NP}[3]{{\jf Nucl. Phys.\jt} \VPY{#1}{#2}{#3}}
\newcommand{\PHYREP}[3]{{\jf Phys. Rept.\jt} \VPY{#1}{#2}{#3}}
\newcommand{\PL}[3]{{\jf Phys. Lett.\jt} \VPY{#1}{#2}{#3}}
\newcommand{\PRD}[3]{{\jf Phys. Rev. D\jt} \VPY{#1}{#2}{#3}}
\newcommand{\PRL}[3]{{\jf Phys. Rev. Lett.\jt} \VPY{#1}{#2}{#3}}
\newcommand{\PTRSLA}[3]{{\jf Phil. Trans. R. Soc. Lond.\jt\ A}
\VPY{#1}{#2}{#3}}
\newcommand{\ZhETF}[3]{{\jf Zh. Eksp. Teor. Fiz.\jt} \VPY{#1}{#2}{#3}}


\begin{thebibliography}{99}

\bibitem{Guth1}
A\.H.~Guth, \PRD{23}{347--356}{1981}.
% The Inflationary Universe:  A Possible Solution to the Horizon
%    and Flatness Problems

\bibitem{Starobinsky}
A\.A.~Starobinsky, \ZhETF{30}{719}{1979} [\JETP{30}{682}{1979}];
A\.A.~Starobinsky, \PL{91B}{99--102}{1980}.
% A New Type of Isotropic Cosmological Models without Singularity
%   (yes, that really is ``Models''; reprinted in Abbott & Pi)

\bibitem{Coleman}
S.~Coleman, \PRD{15}{2929--2936}{1977} [see errata
\VPY{16}{1248}{1977}];
% The Fate of the False Vacuum. 1. Semiclassical Theory
C\.G.~Callan and S.~Coleman,
\PRD{16}{1762--1768}{1977}.
% The Fate of the False Vacuum. 2. First Quantum Corrections

\bibitem{HMS}
S\.W.~Hawking, I\.G.~Moss, and J\.M.~Stewart,
\PRD{26}{2681--2693}{1982}.
% Bubble Collisions in the Very Early Universe

\bibitem{GuthWeinberg}
A\.H.~Guth and E\.J.~Weinberg, \NP{B212}{321--364}{1983}.
% Could the Universe Have Recovered from a Slow First Order Phase
%    Transition?

\bibitem{Linde1}
A\.D.~Linde, \PL{108B}{389--393}{1982}.
% A New Inflationary Universe Scenario: A Possible Solution of the
% Horizon, Flatness, Homogeneity, Isotropy and Primordial Monopole
% Problems

\bibitem{Albrecht-Steinhardt1}
A.~Albrecht and P\.J.~Steinhardt, \PRL{48}{1220--1223}{1982}.
% Cosmology for Grand Unified Theories with Radiatively Induced
%    Symmetry Breaking

\bibitem{chaotic}
A\.D.~Linde, \ZhETF{38}{149--151}{1983} [\JETP{38}{176--179}{1983}];
% Chaotic inflating universe
A\.D.~Linde, \PL{129B}{177--181}{1983}.
% Chaotic inflation

\bibitem{hyb1} 
A\.D.~Linde, \PL{B259}{38--47}{1991}.
% Axions in Inflationary Cosmology

\bibitem{hyb2}
A\.R.~Liddle and D\.H.~Lyth, \PHYREP{231}{1--105}{1993},
astro-ph/9303019.
% The Cold Dark Matter Density Perturbation

\bibitem{hyb3}
A\.D.~Linde, \PRD{49}{748--754}{1994}, astro-ph/9307002.
% Hybrid Inflation

\bibitem{hyb4}
E\.J.~Copeland, A\.R.~Liddle, D\.H.~Lyth, E\.D.~Stewart, and
D.~Wands, \PRD{49}{6410-6433}{1994}, astro-ph/9401011.
% False Vacuum Inflation with Einstein Gravity

\bibitem{hyb5}
E.~Stewart, \PL{B345}{414--415}{1995}, astro-ph/9407040.
% Mutated Hybrid Inflation

\bibitem{RSG}
L.~Randall, M.~Solja\v{c}i\'{c}, and A\.H.~Guth,
\NP{B472}{377--408}{1996}, hep-ph/9512439; also hep-ph/9601296.
% Supernatural Inflation: Inflation from Supersymmetry
%   with No (Very) Small Parameters
% Supernatural Inflation (hep-ph/9601296).

\bibitem{Jensen-Stein-Schabes}
L\.G. Jensen and J\.A. Stein-Schabes, \PRD{35}{1146--1150}{1987},
and references therein.
% Is inflation natural?
% Lars Gerhard Jensen and Jaime A. Stein-Schabes 

\bibitem{Starobinsky2}
A\.A.~Starobinsky, \PL{117B}{175--178}{1982}.
% Dynamics of phase transition in the new inflationary universe
% scenario and generation of perturbations

\bibitem{GuthPi}
A\.H.~Guth and S.-Y.~Pi, \PRL{49}{1110--1113}{1982}.
% Fluctuations in the new inflationary universe

\bibitem{Hawking1}
S\.W.~Hawking, \PL{115B}{295--297}{1982}.
% The development of irregularities in a single bubble
% inflationary universe

\bibitem{BST}
J\.M.~Bardeen, P\.J.~Steinhardt, and M\.S.~Turner,
\PRD{28}{679--693}{1983}.
% Spontaneous creation of almost scale-free density perturbations
% in an inflationary universe

% For the current line width settings, the spacing TeX gives is
% ridiculous if one doesn't lower \hyphenpenalty so that
% ``Brandenberger'' can be hyphenated:
{\hyphenpenalty=100
\bibitem{BFM}
For a modern review, see
V\.F.~Mukhanov, H\.A.~Feldman, and R\.H.~Brandenberger,
\PHYREP{215}{203--333}{1992}.\par}
% Theory of Cosmological Perturbations

\bibitem{Guth-RS}
A\.H.~Guth, \PTRSLA{307}{141--148}{1982}.
% Phase transitions in the embryo universe

\bibitem{dicke}
R\.H.~Dicke and P\.J\.E.~Peebles,  in {\bf General
Relativity: An Einstein Centenary Survey}, eds: S\.W.~Hawking and
W.~Israel (Cambridge University Press, 1979).

\bibitem{preskill}
J\.P.~Preskill, \PRL{43}{1365--1368}{1979}.
% Cosmological production of superheavy magnetic monopoles

\bibitem{bond-jaffe}
J\.R.~Bond and A\.H.~Jaffe, talk given at Royal Society Meeting
on {\bf The Development of Large Scale Structure in the
Universe,} London, England, 25-26 Mar 1998, submitted to {\jf
Phil. Trans. Roy. Soc. Lond. A}, astro-ph/9809043.

\bibitem{steinhardt-nuffield}
P\.J. Steinhardt, in {\bf The Very Early Universe}, Proceedings
of the Nuffield Workshop, Cambridge, 21 June -- 9 July, 1982,
eds: G\.W.~Gibbons, S\.W.~Hawking, and S\.T\.C.~Siklos (Cambridge
University Press, 1983), pp. 251--266.
% Natural Inflation

\bibitem{vilenkin-eternal}
A.~Vilenkin, \PRD{27}{2848--2855}{1983}.
% The Birth of Inflationary Universes.

\bibitem{guth-pi2}
A\.H.~Guth and S.-Y.~Pi, \PRD{32}{1899--1920}{1985}.
% Quantum Mechanics of the Scalar Field in the New Inflationary
%    Universe

\bibitem{coleman-deluccia}
S.~Coleman \& F.~De~Luccia, \PRD{21}{3305--3315}{1980}.
% Gravitational effects on and of vacuum decay

\bibitem{vvw}
V.~Vanchurin, A.~Vilenkin, \& S.~Winitzki, gr-qc/9905097.
% Predictability crisis in inflationary cosmology and its
%   resolution

\bibitem{aryal-vilenkin}
M.~Aryal and A.~Vilenkin, \PL{199B}{351--357}{1987}.
% The Fractal Dimension of Inflationary Universe.
% Mukunda Aryal

\bibitem{linde-eternal}
A\.D.~Linde, \MPL{A1}{81}{1986};
% Eternal Chaotic Inflation
A\.D.~Linde, \PL{175B}{395--400}{1986};
% Eternally Existing Selfreproducing Chaotic Inflationary Universe
A\.S.~Goncharov, A\.D.~Linde, and V\.F.~Mukhanov,
\IJMODPHYS{A2}{561--591}{1987}.
% The Global Structure of the Inflationary Universe.

\bibitem{random-vil-ford}
A.~Vilenkin and L\.H.~Ford, \PRD{26}{1231--1241}{1982}.
% Gravitational Effects Upon Cosmological Phase Transitions.

\bibitem{random-linde}
A\.D.~Linde, \PL{B116}{335}{1982}.
% Scalar Field Fluctuations in Expanding Universe and the New
%    Inflationary Universe Scenario.

\bibitem{random-starobinsky}
A.~Starobinsky, in {\bf Field Theory, Quantum Gravity and
Strings}, eds: H.J. de Vega \& N. S\'anchez, {\jf Lecture Notes
in Physics\jt} (Springer Verlag) Vol.~246, pp.~107--126 (1986).

\bibitem{linde-book}
See for example A\.D.~Linde, {\bf Particle Physics and
Inflationary Cosmology} (Harwood Academic Publishers, Chur,
Switzerland, 1990) Secs.~1.7--1.8.

\bibitem{hartle-hawking}
J\.B.~Hartle \& S\.W.~Hawking, \PRD{28}{2960--2975}{1983}.
% Wave Function of the Universe

\bibitem{tunnel-vilenkin}
A.~Vilenkin, \PRD{30}{509--511}{1984};
% Quantum Creation of Universes
A.~Vilenkin, \PRD{33}{3560--3569}{1986};
% Boundary Conditions in Quantum Cosmology
A.~Vilenkin, gr-qc/9812027, to be published in {\bf Proceedings
of COSMO 98}, Monterey, CA, 15-20 November, 1998.
% The Quantum Cosmology Debate

\bibitem{tunnel-linde}
A\.D. Linde, \NC{39}{401--405}{1984};
% Quantum Creation of the Inflationary Universe
A\.D. Linde, \PRD{58}{083514}{1998}, gr-qc/9802038.
% Quantum Creation of an Open Inflationary Universe

\bibitem{hawking-turok}
S\.W.~Hawking \& N\.G.~Turok, \PL{B425}{25--32}{1998}, hep-th/9802030.
% Open Inflation Without False Vacua.

\bibitem{LLM}
A.~Linde, D.~Linde, \& A.~Mezhlumian, \PRD{49}{1783--1826}{1994},
gr-qc/9306035.
% From the Big Bang Theory to the Theory of a Stationary
%   Universe

\bibitem{GBLinde}
J.~Garcia-Bellido \& A.~Linde, \PRD{51}{429--443}{1995},
hep-th/9408023.
% Stationarity of Inflation and Predictions of Quantum Cosmology

\bibitem{borde-vilenkin}
A.~Borde \& A.~Vilenkin, \PRL{72}{3305--3309}{1994}, gr-qc/9312022.
% Eternal inflation and the initial singularity

\bibitem{borde-vilenkin2}
A.~Borde \& A.~Vilenkin, \PRD{56}{717--723}{1997}, gr-qc/9702019.
% Violations of the Weak Energy Condition in Inflating
%   Space-Times.

\bibitem{center-world}
A.~Linde, D.~Linde, \& A.~Mezhlumian, \PL{B345}{203--210}{1995},
hep-th/9411111.

\bibitem{vilenkin-proposal}
A.~Vilenkin, \PRL{81}{5501--5504}{1998}, hep-th/9806185.
% Unambiguous Probabilities in an Eternally Inflating Universe

\end{thebibliography}
\end{document}